# Coupling-enhanced Dual ITO Layer Electro-absorption Modulator in Silicon Photonics

MOHAMMAD H. TAHERSIMA[1,*], ZHIZHEN MA[1,*], YALIANG GUI[1], SHUAI SUN[1], HAO WANG, RUBAB AMIN[1], HAMED DALIR[2], RAY CHEN[3], MARIO MISCUGLIO[1], VOLKER J. SORGER[1,+]

[1]George Washington University, 800 22nd Street NW, Washington, DC 20052, USA

[2]Omega Optics, Inc. 8500 Shoal Creek Blvd., Bldg. 4, Suite 200, Austin, Texas 78757, USA

[3]Department of Electrical and Computer Engineering, University of Texas at Austin, Austin, Texas 78712, USA

*Equal author contribution.

+Corresponding author: sorger@gwu.edu

# Abstract

Electro-optic signal modulation provides a key functionality in modern technology and information networks. Photonic integration has enabled not only miniaturizing photonic components, but also provided performance improvements due to co-design addressing both electrical and optical device rules. However, the millimeter-to-centimeter large footprint of many foundry-ready photonic electro-optic modulators significantly limits on-chip scaling density. To address these limitations, here we experimentally demonstrate a coupling-enhanced electro-absorption modulator by heterogeneously integrating a novel dual-gated indium-tin-oxide (ITO) phase-shifting tunable absorber placed at a silicon directional coupler region. Our experimental modulator shows a 2 dB extinction ratio for a just 4 μm short device at 4 volt bias. Since no material nor optical resonances are deployed, this device shows spectrally broadband operation as demonstrated here across the entire C-band. In conclusion we demonstrate a modulator utilizing strong index-change from both real and imaginary part of active material enabling compact and high-performing modulators using semiconductor foundry-near materials.

# Introduction

Integrated electro-optic (EO) modulators perform key applications in tele- and data communication[1], inter-chip and possibly intra (on-chip) photonic interconnects for multicore microprocessors and memory systems[2,3], RF and analog signal processing such as photonic A/D conversion[4,5] and in sensors[6]. Monolithic integration, such as in silicon photonics, enables i) densifying photonic networks compared to discreetly-packaged components, ii) reduced device power-consumption[7], and iii) enables a platform approach for cost- and density-scaling due to synergies arising largely processing Silicon-based components. The weak EO properties of Silicon[8], however, result in order of millimeter-to-centimeter large modulator footprints, and thus impedes large-scale integration strategies, which was a major driver for the chip industry over decades[9]. Recent explorations in using photonic integrated circuits (PIC) for the interconnectivity functions of neural network also point to the importance of densification[10,11].

The performance metrics for modulators[12,13] are high speed (3dB role-off, $f_{3dB}$)[14,15], high sensitivity/energy efficiency (E/bit, dynamic power)[7,16,17], high extinction ratio and compact footprint (ER/unit-length)[15]. In order achieve high-performance modulators, two possible routes can be followed; i) increasing the light-matter interaction (LMI) such as increasing modal overlap, using optical resonances feedback (i.e. cavity Finesse, $\mathcal{F}$), or utilizing a high-group index (slow-light), and ii) utilizing materials with strong optical index tunability[18]. On-chip plasmonic modulators (surface plasmonic or hybrid plasmonic waveguide) were introduced as an efficient LMI enhancement approaches due to the large group index in plasmonic mode, resulting in compact footprint and high-bandwidth devices[15,19], however, the significant ohmic loss from plasmonic effect limits the propagation distance of information encoded with light[20], thus the modulation power efficiency suffers from optical power penalty[21]. On the other hand, silicon photonic micro-cavities with high quality ($Q$) factors were adopted to boost the LMI with photons interacting multiple times with the EO-tuned material proportional to the cavity's $\mathcal{F}$.[22], but the fundamental trade-off between the photon life-time and device temporal response ($f_{3dB}$) constrain the modulator performance, also, cavities with high $Q$ would not operate spectrally broadband, thus requiring thermal resonance tuning to match the desired wavelength from the PIC, further increasing the modulation energy consumption via addition of this static power. To overcome the material limitation of Silicon as EO 'active' material, heterogeneous integration[23] has been introduced as a possible route to continue device scaling[12] while minimizing modulator signaling performance; the idea is to utilize the foundry established Silicon platforms for passive waveguide parts for its low-loss propagation waveguide system, but to deploy other material options with stronger index modulation for active light-manipulation. The weak index tuning of Silicon, is due to a) the low carrier induced plasma dispersion, and b) the relatively low bandgap. A variety of EO-functional materials, such as graphene[24,25], III-V materials[26] and transparent conductive oxides (TCO)[27,28] have been investigated for EO modulation. Here we select Indium-tin-oxide (ITO) as the EO material and integrate an ITO-oxide-ITO capacitive gate-stack atop a silicon photonic waveguide to form an all-photonic optical mode. The choice for ITO is threefold; i) the carrier concentration modulation in ITO can be in the order of ~$10^{20}$ cm$^{-3}$, which results in a refractive tuning in unity order[19,27], ii) Unity-strong optical index modulation ($\Delta n \sim 1$) has been experimentally observed at, or close to, the epsilon-near-zero (ENZ) condition[29,30], iii) While process control of ITO has been challenging due to intrinsic material complexities, we have yet recently demonstrated a holistic (electrical and optical) approach to both precisely and repeatedly control ITO's parameters[31]. iv) ITO belongs to the class of TCO family and is currently used massively by both smart phones and solar-cell industry for touch screens, and transparent low-resistive front-end contact, respectively[32]. Hence it might enter foundry processes sooner than other exotic materials.

With the exceptional refractive index modulation, ITO has been heterogeneously integrated with photonic waveguide to build next generation EO modulators. If using solely the imaginary part index tuning, with optimization of optical mode and material processing condition, despite the material already gated to the ENZ point, photonic electro-absorption (EA) ITO modulator has been reported to have only 0.15 dB/μm extinction ratio (ER)[29], due to limited light-ITO thin film interaction. Recently, phase-shift based ITO modulator has been also demonstrated by a Mach-Zehnder interferometer (MZI) configuration[33], but the modulator performance is limited by fundamental Kramers-Kronig relation, thus having either a high insertion loss (*IL*), or exhibits degraded modulation depth as the loss and phase of the material changes simultaneously With the addition of a photonic crystal cavity design, Wang's group demonstrated a nanocavity-based ITO modulator with ultracompact footprint[34], they also claim that for such device the phase- and absorption modulation of the material contribute coherently to the EO modulation. The concept of using both real and imaginary part index tuning for modulation is indeed interesting, however, here we argue that for nanocavity-based devices, the extra absorption induced from ITO carrier accumulation would reduce the Q of the cavity thus broadening the spectrum response[35]. As a result, the overall extinction performance of such modulator would be degraded by the extra loss accompanying the phase change of the material. Nevertheless, the aforementioned heterogeneously integrated ITO photonic modulators bias (gating) ITO against Silicon, which usually has a low conductivity thus leads to high RC delay. Selective doping of Silicon could reduce the serial resistance for potentially achieving high-speed modulation, but which increase the design and fabrication complexity, and loss.

Here, we experimentally demonstrate a compact coupling-enhanced dual-layer ITO modulator that uses both the real and imaginary parts of ITO's tunable index synergistically to increase EO modulation while not relying on optical or material resonances thus allowing for truly broadband performance. In brief, in the absorptive stage (light OFF) of the modulator, an additional waveguide section ('coupling island') in close proximity to the bus waveguide enables a higher extinction ratio compared to the traditional absorption-only case without sacrificing device footprint by realizing light power diversion (through coupling) away the bus to the island. While the light-ON stage the coupling is minimized and the optical losses are minimal due to the photonic (non-plasmonic) mode and dielectric-like dual-gated ITO/oxide/ITO stack atop both waveguides (bus and island) to change the both the EAM loss and the mode coupling. We experimentally achieve more than 2 dB of ER from a 4 μm short device (0.5 dB/μm), benefiting from both the loss and coupling modulation of the system, and low energy consumption of 0.77 pJ/bit. The nature of the directional coupling and free-carrier absorption from ITO enables a spectrally broadband response, which can be seamlessly integrated with wavelength division multiplexing (WDM) technology for higher link-level data rates. Also, the dual stack ITO configuration eliminates the high serial resistance from Silicon, thus enables a potentially short RC delay.

**Results**

For directional coupler (DC) devices, the study of a ITO based directional switch was first introduced in 2015[36], where a plasmonic mode and ITO/metal stack are placed on the bar waveguide to enhance the light-ITO interaction to realize large index mismatch between bus and bar waveguide to control the coupling between waveguides. Also, unlike other phase change materials such as GeSbTe (GST) that could change from dielectric to metallic phase[37], the low mode overlap of the active region in such a photonic mode (ITO/Silicon waveguide) would not provide a high effective index change for efficient switching (see supplementary information iii). In contrast in this work, we explore a different and, thus far unexplored, modulation scheme by stacking ITO/oxide/ITO layer across both the bus and the coupling-island to alter both the absorption and beating length of such a coupled system (Fig. 1). Without bias, the dual ITO layers are in flat band condition (assuming same doping levels verified in ref[31], thus the ITO layers are in a low-loss state (ITO is dielectric with low carrier concentration). Here, the DC's beating length is much longer than the island length resulting in high transmission at the bus waveguide (Fig. 1a). By capacitively gating the dual ITO layers one achieves a higher linear absorption (loss) compared to the zero bias state due to carrier accumulation. Note, the dual ITO layers operating in a seemingly push-pull configuration appear as there would be a vanishing net index change (one ITO layer carrier accumulation the other ITO depletion). However, this is not true, since ITO's Drude-model shows a non-symmetric index behavior with carrier concentration change (Fig. 3a). In addition, the carrier accumulation greatly reduces the refractive index of ITO (order of unity) atop the waveguides and inside the gap (between both waveguides), thus the optical path between two waveguides is reduced as if the gap between the bus and island waveguide would be shrunk. Since the coupling coefficient is rather sensitive to the gap size between the two waveguides, more optical power is transferred to the island waveguide with electrical bias and a low transmission state from bus waveguide can be observed (Fig. 1b). Here, between the light ON and OFF states, the synergistic usage of (turned-on with bias) absorption and coupling together help to reduce the light intensity in bus waveguide. As an EO modulator, we claim that the real and imaginary part tuning of ITO contribute together to the modulation depth (extinction ratio, ER).

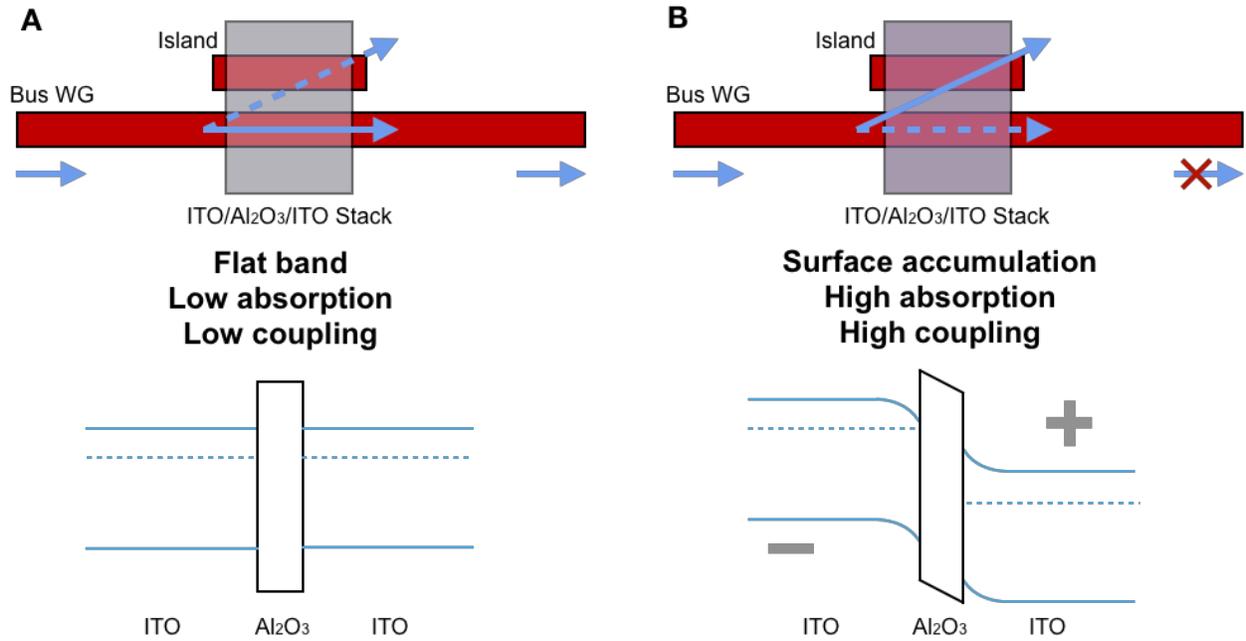

**Figure 1**. Operation principle for a coupling-enhanced ITO modulator, here both the change in ITO absorption and change of coupling coefficient contribute together to the modulation of the transmission on the bus waveguide. A. Without external bias, the top and bottom layer of ITO has an intrinsically low carrier concentration, thus low absorption for the device region. Also, a long beating length prohibit the coupling from bus waveguide to the island waveguide, a high transmission could be expected from the bus waveguide. B. With external gating applied, the bottom layer ITO becomes more absorptive due to carrier accumulation, while the beating length becomes shorter for the coupling region, the transmission is blocked from the bus waveguide.

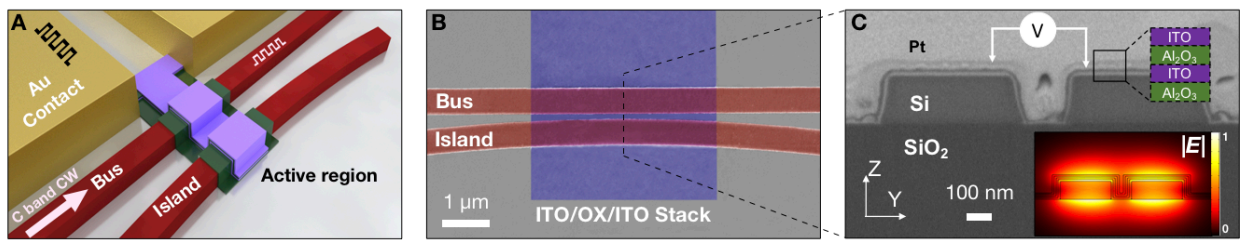

**Figure 2**. **A.** Schematic of coupling-enhanced dual-gated ITO modulator. **B.** Scanning electron microscopy image of fabricated device. **C.** FIBSEM cross-section illustrating the device layers and operating mechanism of the active region. An optical beam is modulated by modifying the coupling coefficient between bus and island waveguides via applying an electrical bias across the double ITO layer capacitor. In the coupling region, two waveguides are spaced by a gap distance g = 200 nm, covered by an active device region (zoom-in: ITO/$Al_2O_3$/ITO). In the active coupling region, the first 10 nm of $Al_2O_3$ layer is directly deposited on top of the Silicon waveguides as a passivation layer using atomic layer deposition (ALD), then the 15 nm inner ITO layer, 10 nm Al2O3 gate oxide and 15 nm outer ITO layer is patterned and deposited on the coupling region consecutively to define the device region, the ITO layers are gated in a capacitively push-pull configuration to modulate the photonic coupler region. Here ITO's optical index changes nonlinearly (i.e. non-symmetrically) with applied bias (see Fig. 3a). This nonlinear behavior is further enhanced by a non-symmetric model overlap of the bottom vs. top ITO-layer (see inset illustrates the corresponding TM mode-field profile from eigenmode analysis).

Our dual-gated ITO modulator consists of a Silicon photonic bus waveguide, and a short (2-8 μm long) coupling-island waveguide separated 200 nm away from the bus. SOI waveguides: width = 500 nm, height = 220 nm is patterned via negative E-beam photoresist (HSQ) followed by dry etching using $SF_6$ and $C_4F_8$, and the dual-gated $Al_2O_3$/ITO/$Al_2O_3$/ITO stack is deposited on top of the coupling region, where the lower oxide serves as electrical isolation. Note, that the short island waveguide is slightly curved away from the bus waveguide to minimize the abrupt effective index change for the propagation, thus minimizing the insertion losses from the SOI waveguide into the device section. Both ITO layers extend beyond the coupling region to facilitate electrical connection via two Ti/Au contact pads, (Fig. 2A). Overall, the 50 nm-tall (vertically) stack is not only on top of the waveguides but also resides inside the gap between, which effects the propagating constant of the TM mode inside the waveguide (Fig. 2C inset). This setup effectively allows voltage-tuning the coupled supermodes index and hence altering the coupling length of the DC. The induced change in the carrier concentration of ITO layers thus tunes the absorption of ITO layers and the coupling factor between two waveguides simultaneously.

The optical permittivity of ITO, for near-infrared (NIR) wavelength, can be described using Drude model as[27,29]:

$$\varepsilon_{ITO}(\omega) = \varepsilon_\infty - \frac{\omega_p^2}{\omega^2 + i\omega\Gamma} \quad (1)$$

where $\omega$ is the angular frequency of light, $\varepsilon_\infty$ is the high frequency permittivity of ITO, $\Gamma$ the damping frequency for the electron collisions inside ITO. And $\omega_p = \sqrt{\frac{N_c e^2}{\varepsilon_0 m^*}}$ is the plasma frequency, depends on the carrier concentration $N_c$, and effective mass of electrons inside ITO film $m^*$. Thus, by changing the carrier concentration $N_c$ in ITO electrically, the optical permittivity is tuned. It has been reported that for ITO deposition, depending on the processing and post-processing conditions, such as the gas mixture, temperature during sputter and post-annealing, the intrinsic carrier concentration could span from $1 \times 10^{19}$ cm$^{-3}$ to $1 \times 10^{21}$ cm$^{-3}$, which greatly affects the optical index for the as-deposited ITO film[38,39]. In our work, we perform Hall-bar and ellipsometry measurement for our ITO thin films on a reference wafer which is processed in the same batch as the device sample, in order to control the ITO thin film optical property and extrapolate the carrier concentration and damping rate of the material. For instance, we measure an initial carrier concentration (without electrical bias) of $1.2 \times 10^{19}$ cm$^{-3}$ for our as-deposited ITO from Hall-bar measurement, which is rather low due to the absence of a post-deposition thermal treatment. Then broadband (193-1690 nm) ellipsometric spectroscopy is performed and fitted with the Lorentz oscillator, Cauchy model and Drude models in order to extract the ITOs film parameters at zero bias giving a damping rate of $9.5 \times 10^{13}$ rad/s, at the lowest mean square error (MSE)[31]. Fabricated devices, (with the 10 nm $Al_2O_3$ gate oxide) show an ITO carrier concentration in the accumulation layer of $5.1 \times 10^{20}$ cm$^{-3}$ for 4 V electrical bias, while the optical property of ITO with carrier concentration is plotted in Fig. 3A, for 1550 nm wavelength in our device the accumulation layer of ITO is tuned to be close to ENZ point (~$6.8 \times 10^{20}$ cm$^{-3}$), achieving nearly unity change in real part of refractive index, and a significant increase in optical loss ($\kappa$) of the ITO film.

Given the slight curvature of the coupling-island relative to the waveguide bus, the effective coupling length is shorter than the physical (ITO/oxide/ITO) stack; for instance, for the 4 μm long active device region this effective coupler length is actually 3.4 μm after normalization to a straight coupler with 200 nm gap size. Without bias both ITO layers exhibit dielectric material property with low loss. Performing a 2D eigenmode analysis, we find the effective index difference (phase mismatch) between the first order and second order mode to be 0.099, with a low linear absorption of $\alpha = 21.89$ cm$^{-1}$ (see supplementary information iv). The DC coupler's beating length in this case can be estimated by

$$L_{beat} = \frac{\lambda}{2(n_{even} - n_{odd})} \approx 7.8 \text{ μm} \quad (2)$$

where the $\lambda = 1.55$ μm is the operating wavelength, yielding an insertion loss of 2.1 dB for this 4 μm long device at on state. With applied bias of 4 V (i.e. inner ITO layer biased to accumulation), the stronger free-carrier absorption from ITO increases the linear absorption of the supermode to be $\alpha = 146$ cm$^{-1}$. Meanwhile, the near-unity index reduction of ITO's real part reduces the effective optical path between the bus and island waveguide, resulting in an increased phase mismatch between the first and second order mode of 0.117, consecutively corresponds to a shorter beating length of 6.6 μm. Thus, more optical energy transfers from the bus to the island waveguide while the increased absorption further simultaneously decreases the transmission of the bus waveguide compared to the ON-state. Thus the increased coupling coefficient from real index modulation and higher free-carrier absorption accompanying the real index change contribute simultaneously to the intensity modulation on the bus waveguide, resulting in an ER = 2 dB for this 4 µm device. Interestingly, we observe a weaker modulation when the outer ITO layer is in accumulation state, which is due to the optical perturbation theory that the outer ITO is further away from the Silicon waveguide core, resulting a smaller ITO mode overlap (Fig. 3B). Also, the modulator performance is measured to be quite uniform across the C-band, because of the non-resonant nature of both the coupling effect and the intra-band free-carrier-based absorption from ITO (Fig. 3C). Lastly, the length dependent study of effective (i.e. bending-corrected) coupling length versus ER performance shows reasonable agreement between the numerical approach and experimental measurement (Fig. 3D). This indicates that indeed the change of linear absorption and coupling coefficient contribute synergistically towards higher ER.

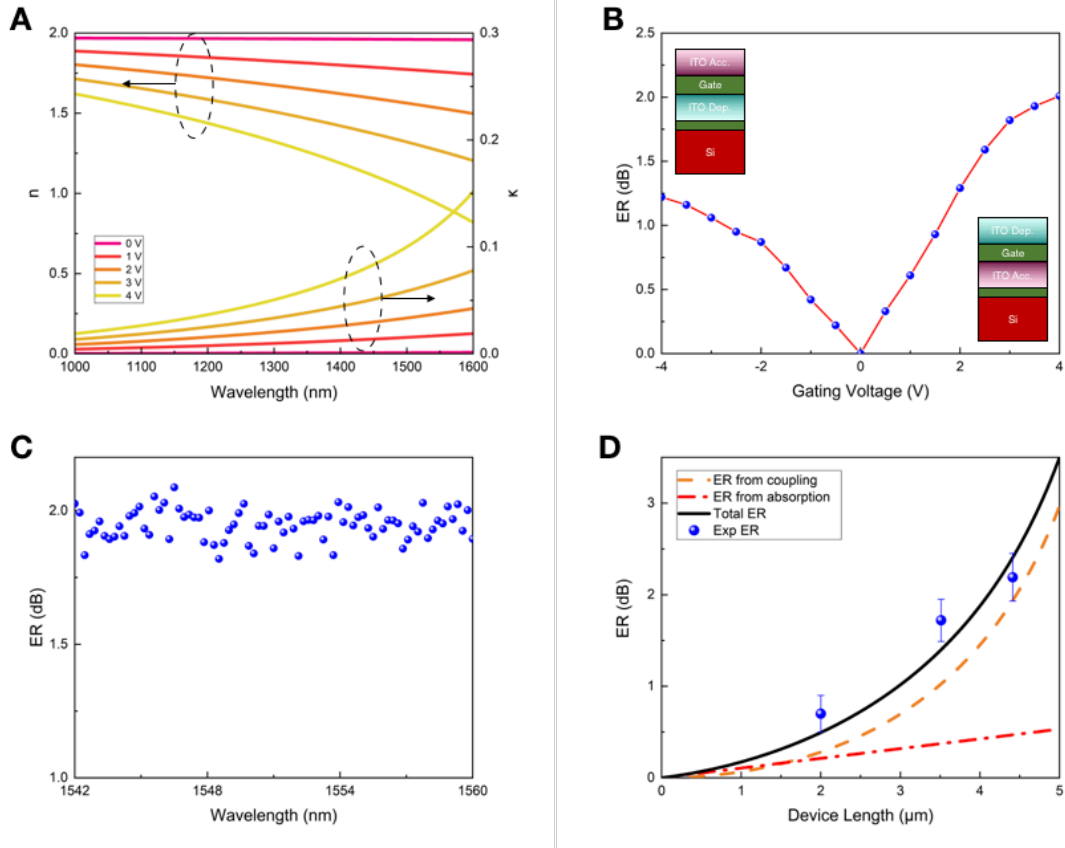

**Figure 3**. A. Experimental verified ITO refractive index data under different carrier concentration, fitted using Drude model. B. Modulation performance for a 4 µm device, note that for outer layer ITO accumulation shows a weaker modulation effect compare to inner ITO accumulation, since it is further away to the waveguide core. C. Broadband performance of the modulator across C band, the limitation is mainly due to the spectrum response of the grating coupler. D. Length dependent measurement for 3 different device length (2 µm, 4 µm and 6 µm long devices have been fabricated and measured, which correspond to an effective coupling device length of 1.95 µm, 3.4 µm and 4.4 µm, respectively.) shows a good agreement with the numerical simulation result, illustrating the contribution of linear absorption and coupling change for the ER.

In addition to the joint contribution from both the real and imaginary index change towards a mutually synergistic modulation operation, the dual-gated ITO layer also supports overcoming the high serial resistance drawbacks of devices where ITO is biased against Silicon[33,34]. Providing some estimation to this effect here, for our 4 μm device, the metal contacts are spaced 6 μm away from the active region, thus the serial resistance is 1.8 KΩ using our measured resistivity of deposited ITO thin film of $1.8 \times 10^{-5}$ Ω·m from Hall bar measurement, which is even a few times lower than those reported for doped Silicon waveguide as the contact[34]. The estimated RC limited bandwidth of our device is 458 MHz, with the capacitance of 193 fF. However, with proper optimization such as reducing contact spacing of the device region both the serial resistance and capacitance could be reduced to 525 Ω and 56 fF, respectively, which corresponds to a RC bandwidth of 54 GHz, and modulation energy consumption of 200 fJ/bit, while keeping a minimum distance of half wavelength (750 nm) to avoid generating plasmonic effect. Overall, with 2 dB modulation from a 4 μm device, such a modulation of 0.5dB/μm is rather high compared to other device schemes, and does not even rely on resonances (spectrally broadband). With an IL = 2 dB, the ER/IL-ratio is about unity; while this is not outstanding compared to lithium-niobate[40] or even Silicon[21] modulators, these devices are 3-4 orders of magnitude more compact. Also, by engineering the device dimension, for instance, reduce the gap size below 200 nm, or increase the device length to reach two times of beating length at ON state, would further enhance the ER/IL performance of the device while keep the device footprint still in sub-15 μm range, which is much shorter than a linear absorption only ITO modulator[29].

**Conclusion**

In summary, we introduce and experimentally demonstrate a novel coupling-enhanced dual-gated ITO modulator heterogeneously integrated at the coupling regime between a silicon waveguide bus and a short coupling island. We show that for our coupling-enhanced electro-absorption modulator, both the real and imaginary index modulation of this ITO/oxide/ITO stack synergistically contribute towards the modulation. This way we achieve a 2 dB extinction ratio, ER, modulation while being 4 μm compact (0.5 dB/μm) with an insertion loss, IL, of 2 dB (ER/IL = 1) for 4 volts of applied bias. We verify a flat spectral response across the C-band frequencies since neither optical nor material resonances are needed in this device configuration. The design features careful process control to enable an ITO material away from the ENZ-point at light ON-state to reduce losses while demonstrating a near unity ITO EO index change under modulation. Taken together, the heterogeneous integration of an emerging electro-optic material, ITO, which is commonly used in the semiconductor industry, offers positive device-synergies demonstrating a compact coupling-enhanced modulator on a Silicon photonic platform.

## Acknowledgement

VS is funded by AFOSR (FA9550-17-1-0377) and ARO (W911NF-16-2-0194) and HD by NASA STTR, Phase I (80NSSC18P2146).